%% file: main.tex
\documentclass[12pt, a4paper]{article}
\usepackage[margin=1in]{geometry}
\usepackage{booktabs, longtable, adjustbox, rotating}
\usepackage{multirow, tabularx}

\usepackage{etoolbox}

\usepackage[T2A]{fontenc}
\usepackage[utf8]{inputenc}
\usepackage[russian, english]{babel}

\usepackage{comment}
\usepackage{tikz, bm, amsmath}
\usepackage{soul}
\usepackage{bbm}
\usepackage{glossaries}
\usepackage{graphicx}               

\graphicspath{{./assets/}} 

\usepackage{fancyhdr}
\newglossaryentry{impactv}{
	name={Impact v. influence},
	description={``Causal impact'' or ``impact'' refers to the causal influence one outlet has on another in quotation usage, and the term ``influence'' refers to the general concept of effects one outlet may have on another}
}

\newglossaryentry{impactnetwork}{
	name={Impact network},
	description={Estimated causal impact between media outlets represented as a directed, weighted graph where nodes are outlets and connecting edges are the causal impact}
}

\newglossaryentry{influencenetwork}{
	name={Influence network},
	description={Channels of potential influence between media outlets represented as a directed, weighted graph where nodes are outlets and connecting edges are the saliency-weighted volume of historical quote following}
}

\newglossaryentry{mediaorientation}{
	name={Media outlet orientation},
	description={Classification of an outlet's political orientation based on qualitative assessment of its published content along with ownership analysis}
}

\newglossaryentry{percentcausedby}{
	name={Percentage of quote following caused by ...},
	description={Normalized causal impact by dividing by the maximum potential impact. A normalized impact $x$ of outlet $i$ on outlet $j$ can be thought of as ``of the times $j$ uses the same quotes as $i$, $x\%$ of the time it was caused by exposure to $i$''}
}

\newglossaryentry{prous}{
	name={Pro-U.S. vs. pro-Russia},
	description={To demonstrate our approach, we analyze Russian state-controlled media and how it shapes media sentiment favorable towards its competition with the U.S. on various topics. The competing sentiments in our analysis are pro-U.S. versus pro-Russia}
}

\newglossaryentry{saliency}{
	name={Saliency-weighted potential quote influence},
	description={Potential influence of one outlet on another measured in potential quote following weighted by saliency criteria described in Section ``Method.Saliency-Weighted Potential Quote Influence''}
}

\newglossaryentry{sentiment}{
	name={Sentiment},
	description={The political position the article takes in using the quotation}
}

\newglossaryentry{slant}{
	name={Slant},
	description={``Impact slant'' or ``slant'' refers to the difference between an outlet's causal impact in competing sentiments. For example, if an outlet has larger pro-U.S. than pro-Russia impact, it has a pro-U.S. slant}
}

\newglossaryentry{totalimpact}{
	name={Total impact},
	description={The sum of causal impacts of one outlet on others over all sentiments}
}
\makeglossaries

\title{\Large Exposing the Obscured Influence of State-Controlled Media: \\A Causal Estimation of Influence Between Media Outlets Via Quotation Propagation}
\author{\small \hspace{-1.5em} Joseph Schlessinger$^{1,3}$, Richard Bennet$^2$, Jacob Coakwell$^2$, Steven T. Smith$^1$, \& Edward K. Kao$^1$}
\date{\small
	$^1$MIT Lincoln Laboratory\\%
	$^2$Novetta, LLC\\%
	$^3$MIT Institute for Data, Systems, and Society
}

\begin{document}
	\maketitle

\begin{abstract} \normalsize
\noindent This study quantifies influence between media outlets by applying a novel methodology that uses causal effect estimation on networks and transformer language models. We demonstrate the obscured influence of state-controlled outlets over other outlets, regardless of orientation, by analyzing a large data set of quotations from over 100 thousand articles published by the most prominent European and Russian traditional media outlets, appearing between May 2018 and October 2019. The analysis maps out the network structure of influence with news wire services serving as prominent bridges that connect outlets in different geo-political spheres. Overall, this approach demonstrates capabilities to identify and quantify the channels of influence in intermedia agenda setting over specific topics. 
\end{abstract}

\vfill
\noindent\rule{3.7cm}{0.5pt}\\
\small
\noindent Corresponding authors: Edward Kao - edward.kao@ll.mit.edu
 
\noindent\hspace{3.95cm} Richard Bennet - rbennet@novetta.com\\\\

\noindent \scriptsize
DISTRIBUTION STATEMENT A. Approved for public release. Distribution is unlimited.
This material is based upon work supported by the Under Secretary of Defense for Research and Engineering under Air Force Contract No. FA8702-15-D-0001. Any opinions, findings, conclusions or recommendations expressed in this material are those of the author(s) and do not necessarily reflect the views of the Under Secretary of Defense for Research and Engineering.\\
\copyright \ 2021 Massachusetts Institute of Technology.\\
Delivered to the U.S. Government with Unlimited Rights, as defined in DFARS Part 252.227-7013 or 7014 (Feb 2014).\\ Notwithstanding any copyright notice, U.S. Government rights in this work are defined by DFARS 252.227-7013 or DFARS 252.227-7014
\clearpage \normalsize

\section*{Introduction}
The influence that news sources have on each other in setting the agenda for news coverage shapes the opinions and choices of the public. (McCombs, 2002) The ability to quantify media outlet relationships becomes especially important as nation-state actors fund and control international media outlets, selecting topics and deciding how those topics are covered in furtherance of a political agenda. (McCombs and Shaw, 1972; Field et al, 2018) Though much attention has been given to government actors’ propaganda efforts, particularly in the case of Russia (Van Herpen, 2015; Paul and Matthews, 2016; Jones, 2002), the often-subtle ways in which state-affiliated outlets set the agenda for other mainstream news coverage remains obscured. General news consumers and those concerned with media literacy and combating hostile disinformation may lack understanding the impact that government-run media has on broader coverage of important issues.

Media outlets—including print, television, radio, and internet—are not isolated, unidirectional, broadcast mechanisms to their reading public, but also influence each other within the media ecosystem. An influential outlet can affect how other outlets and the broader information environment treat a particular topic, determining the amount of coverage a topic receives, the sentiment in subsequent coverage, and the extent to which coverage is sustained over time.  

Thus, an approach that measures the comparative influence of specific outlets over others can quantify how agendas are pushed across this information ecosystem and ultimately to the general public. A quantifiable metric of inter-outlet influence and narrative propagation among outlets also stands to inform the assessment of strategic communications and influence operations, as we can track the relative effectiveness of influence efforts by quantifying the extent to which other outlets are reporting within the agenda set by state-controlled media.

We propose a novel quantitative approach to measure the effects in intermedia agenda setting and framing, a topic that has been previously studied (McCombs and Shaw, 1972). Our approach is also novel in its ability to measure the obscured state-controlled media influence on the overall media landscape. The approach integrates two novel methods that offer evidence of inter-outlet influence in traditional media, based on shared quotations within articles by different outlets. First, we implement a new method to automate quotations matching by mapping them to a semantic space using the state-of-the-art SBERT transformer language model (Reimers \& Gurevych, 2019) then matching them using density-based clustering. Matched quotations reveal potential quote following behavior between outlets over time, which are then used to construct a network of potential influence between outlets. Second, we apply a novel method of influence estimation that attributes impact by accounting for narrative propagation over the network using a causal inference framework developed by Kao and Rubin (Kao, 2017; Imbens \& Rubin, 2015). This causal framework has been demonstrated on social media networks (Smith et al., 2018, 2021) to estimate influence as the difference in outcomes from network exposures. Intuitively, a similar exposure-based approach makes sense for understanding the relationships between media outlets and how they cover topics. With the European and Russian traditional media data set we use, our approach offers a metric of the impact that Russian state-controlled media have on other outlets, showing the extent to which the Russian state’s framing of topics is spreading in the European media environment.  

For this analysis, we examine general influence patterns across a wide variety of geopolitical topics in European media, as well as a case study on a particular topic we named “Nuclear Treaty” that includes coverage of topics involving Russia-U.S. non-proliferation agreements and statements about observing or violating nuclear treaties such as the Intermediate-Range Nuclear Forces (INF) Treaty. In late 2018 and early 2019, we observed Russian state-aligned outlets voicing a consistent messaging narrative on the topic of nuclear treaties: the United States is breaking existing treaties without cause. Russian officials made such statements, and Russian state-affiliated outlets propagated these statements by quoting the officials prolifically to create a narrative of the United States’ destabilizing behavior (“US was Preparing to Produce Missiles Banned Under INF Over Past 2 Years – MoD”, 2019; “US has been Violating INF Treaty for Years, says Russian Defense Ministry Spokesman”, 2019; “Putin says US Exit from Arms Reduction Treaties Undermines Global Security”, 2019;  “US Worked on INF Treaty-banned Missile for Several Years, says Putin”, 2019;  ``\foreignlanguage{russian}{США приступили к подготовке к производству запрещенных ракет 2 года назад} [US Began Preparation for Producing Banned Rockets 2 Years ago]”, 2019;  “США начали готовиться к производству запрещенных ДРСМД ракет два года назад [US Began Preparation for Producing INF-banned Missiles Two Years ago]”, 2019). Conversely, the U.S. position was that Russia designed and deployed weapons that broke the treaty in the previous several years and that the INF was out of date and irrelevant for current security challenges  (Simons \& Marsons, 2019; “US, Russia Rip up Cold War-era INF Missile Treaty”, 2019). Our network-based approach to estimating influence allows us to understand how certain outlets influence other outlets, thereby propagating narratives voiced in both positions.

Applied to European and Russian media, our approach reveals the implicit ways that Russian state-controlled outlets influence other outlets. Though the state-controlled media exercise no direct authority over other outlets, they exert influence over which topics to report on and whom to quote on those topics. (Field et al., 2018; Van Herpen, 2015)

When an outlet features a quotation within an article, a decision is made by the authors and editors, namely, to feature this particular voice as a means to elucidate details of the article’s topic, or relevant opinions that provide context and insight. These quotations form the syntactical unit of the subsequent news coverage that readers consume (Krippendorf, 1980). The authors have extracted quotations—to include paraphrased quotes—as a means to validly separate discrete units of content. Quotations are particularly well suited as valid syntactical units of traditional media, with limited distortion when separated from the article’s context (see Riff, Lacy, \& Fico, 2005, on the validity of syntactical units). Examining the decisions of whom to quote highlights how outlets influence what issues news consumers pay attention to and what positions they consider, a process referred to as \textit{agenda setting} (McCombs \& Shaw, 1972). For instance, if a media outlet decided to predominately quote U.S. officials regarding the INF Treaty rather than Russian officials, that outlet would present a very different account to its audience than an outlet that relied solely on Russian official sourcing. In sum, the decisions made of whom to quote and how to frame those quotations form the core components of an outlet’s coverage of a news item or topic. In topics with competing narratives like the ‘Nuclear Treaty’, the choice of whom to quote, which quotations to highlight, and how many voices from each side of an argument to include will determine the media coverage to which the reader is exposed, thus justifying quotations as the basic unit of analysis in outlet coverage leaning and influence.

Intermedia agenda setting expands the concept of agenda setting beyond the media-to-audience relationship to analyze the media-to-media relationship before the audience even comes into play. We propose a novel approach to intermedia agenda setting research that focuses on the dissemination of quotes from one media outlet to another. When a quotation is featured in one outlet and then repeated in subsequent outlets, the first outlet may have impacted the coverage of the second. This conjecture establishes a mappable relationship through which one outlet may influence another. This assumption often holds true in practice and falls in line with the cascading network activation model (Entman, 2012; Entman \& Usher, 2018). The cascading network activation model is described as one where “media outlets…create cascades of information among themselves, from, for example, more prestigious to less prestigious outlets,” (Roulet \& Clemente, 2018). For instance, news wires such as \textit{Reuters}, \textit{Agence France Presse}, and \textit{TASS }have greater resources and access to sources to support their reporting. Often other outlets with fewer personnel and connections cite such wire services in their reporting. Though it is possible that two outlets will share a quotation if both had journalists present at the same press conference (i.e. their quotations, though shared, were independent of each other’s coverage), mapping the number of shared quotations between outlets while considering temporal differences in the quotations’ appearance provides a good proxy for establishing the relationship between outlets. This approach is particularly appropriate given the implausibility of determining how and where each outlet obtains the quotations it publishes.

Quotes are a suitable unit of analysis from a practical standpoint, as well. The diffusion of quotes can be reliably and automatically detected. While an outlet may excerpt or otherwise change a quote, the quote remains recognizable. Previous work in this area has relied mainly on article topics as the basic unit of analysis. (Guo \& Vargo, 2020) use Granger-causality of news article “themes” to quantify influence between countries. (Stern et al., 2020) take a similar approach, but perform their analysis at the outlet level using Granger-causality of news article topics and sentiments to construct influence networks. We improve upon this work by using quote sharing as our instrument for influence, which we argue is a more plausible signal of inter-media influence than precedence in article topics and sentiments. Moreover, we apply a novel method to quantify the causal impact of each outlet. For the rest of the paper, “causal impact” or “impact” refer to the causal influence one outlet has on another in quotation usage, and the term “influence” refers to the general concept of effects one outlet may have on another. 

\subsection*{Core Contributions}
This paper describes and demonstrates a novel data-driven and quantitative approach to measure the causal impact between media outlets as they propagate quotations with specific sentiments on the topic of interest. Beyond the initial data curation process, this approach is automated. The approach matches quotations using the state-of-the-art SBERT transformer language model, constructs the network of potential influence between outlets, and estimates actual impact with a novel causal inference framework on networks.

The following quantitative results and core findings are obtained by demonstrating the approach on a data set over a wide selection of the most prominent European and Russian traditional media outlets, containing 618,328 quotations from 123,396 articles published between May 2018 and October 2019.

Russian state-controlled media have an average causal impact on 24.2\% of quote following by other Russian outlets, revealing a significant level of obscured influence where Russian state media set the boundaries (topics, sources, etc.) within which other outlets report, regardless of how pro-Russian these outlets are. This is evident in the independent outlet \textit{Radio Echo Moscow} having 31.3\% of its quote following impacted by Russian state media. Furthermore, state-affiliated outlets in Europe are more influenced by Russian state-run outlets than are outlets independent from the state, as seen in 11 out of the 12 countries in our analysis. 

Evident in the network community structure of the estimated causal impact between outlets, outlets typically influence others within their own geo-political media ecosystem. Exceptions are the wire service outlets that exert impact across multiple media ecosystems. Discovered as the notable bridges of the estimated causal impact network, these wire outlets include \textit{Sputnik}, \textit{TASS}, \textit{Reuters News}, and \textit{Agence France Presse}. Within the Russian ecosystem, state-agenda and state-controlled outlets repeat the same quotes, creating an echo chamber. Moreover, our analysis shows that outlets selectively follow quotes; Western outlets follow the pro-U.S. quotes from Russian outlets, while Russian outlets follow the pro-Russia quotes from Western outlets. 

Comparing the reach of two specific state-sponsored media organizations, \textit{Sputnik} has a greater impact on European media than does \textit{RFE/RL}, with an average causal impact on 21.9\% of quote following versus 16.6\%. However, \textit{RFE/RL} shows an advantage in the media environments of 7 out of 21 countries in our analysis. For a more detailed understanding of the media influence landscape, we can apply our quantitative analysis to specific topics such as the end of the Intermediate-Range Nuclear Forces (INF) Treaty in 2019, and identify the “bellwether” outlets that most strongly impact its coverage and sentiment, as well as their most influential quotations. 

\section*{Method}

\begin{figure}
	\centering
	\includegraphics[width=.75\linewidth]{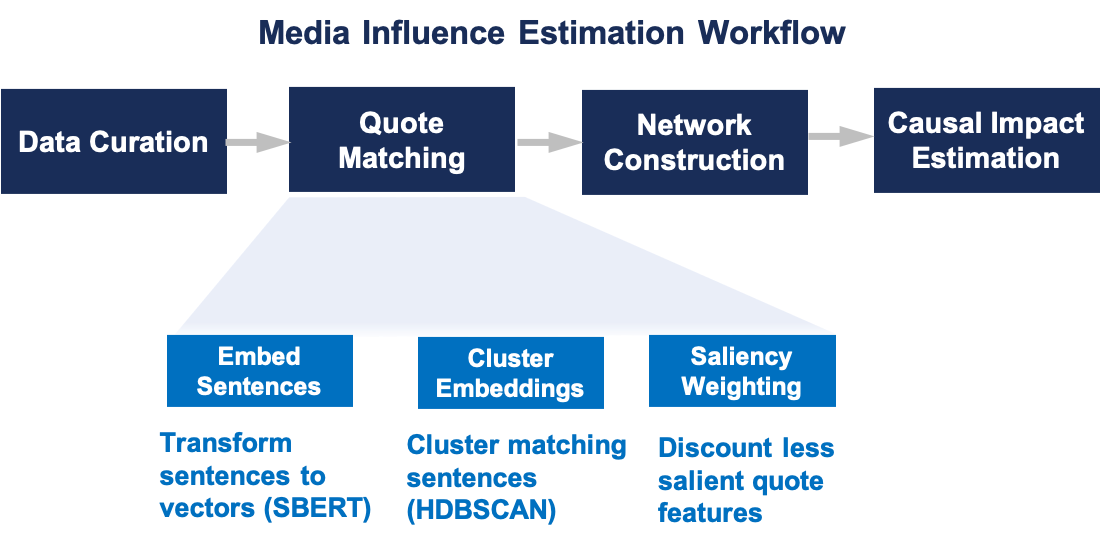}
	\caption{Data-driven methodology for semi-automated media influence estimation}
	\label{fig:block_diagram}
\end{figure}

Our novel data-driven and quantitative approach measures the causal impact between media outlets as they propagate quotations with specific sentiments on the topic of interest. Beyond the initial data curation process, this approach is automated with state-of-the-art text analytics and novel causal inference on networks. Our methodology has four main components, as shown in Figure \ref{fig:block_diagram}. First, we collect a dataset of quotes pulled from articles from prominent European digital media outlets and hand-label them with context-based sentiment and topic. Then, we extract matching quotes from the dataset using a novel and automated method by mapping each quote to a semantic space using the state-of-the-art transformer model SBERT and then identifying matching quotes through clustering with HDBSCAN. Matching quotes are weighted by salience, such that quotes with better signal of influence are weighted more than those with lower signals. Next, we construct an influence network using each media outlet as a node and connecting edges based on historical \textit{quote following} (when an outlet publishes a quote identical to one from another outlet). Finally, we apply a novel network causal inference method to quantify causal impact between outlets.

\subsection*{Data Collection}
For our corpus we use data collected from a wide selection of the most prominent European and Russian traditional media outlets based on third-party services like Alexa.com and SimilarWeb that track the popularity of webpages. The genre of the corpus is geopolitical news coverage. Content was aggregated using standardized search strings filtering for diplomatic, informational, military, and economic content. The data set draws more heavily from countries in eastern Europe than Western Europe but includes the most prominent outlets in 24 countries. Though news consumption preferences vary considerably within every country in Europe, the dataset has selected the most prominent outlets in select countries to approximate the news coverage of geopolitical issues that an average European is exposed to on a regular basis. As such, we have included a small number of U.S.-based outlets to capture the borderless nature of media consumption in Europe. For those outlets based in Russia, the data set includes media affiliated with the Russian government as well as media that is not.

To understand the underlying content and sentiments in each news article, we extract all quotations and label each quotation with who the quoted speaker is, a label of the topic based on a discrete set of geopolitical issues, and a label of the sentiment of that quote toward a particular entity. For example, a quote might be labeled with \textit{speaker} Vladimir Putin, \textit{topic} “discussing Economic Sanctions,” and \textit{sentiment-to-entity} ``negative to U.S.'' Topics and sentiment (positive, negative, neutral) are assigned by a human coder in accordance with guidelines defined in a codebook and are validated by an auditing coder. The coders are subject matter experts with relevant lingual-cultural background. Demonstrating our approach to measure the impact of state-controlled outlets on the overall media landscape, we analyze Russian state-controlled media as the example and how it shapes media sentiment favorable towards its competition with the U.S. on various topics (e.g. the end of the Intermediate-Range Nuclear Forces (INF) Treaty in 2019). Given the focus within our collected data corpus on strategic competition between the United States and Russia, we interpret quotations that are positive toward one of these nations as negative toward the other, and vice-versa. Thus, for our analysis, sentiment is represented as a spectrum, with pro-U.S. on one side and pro-Russia on the other.

We also append information about the date of the article, the country of media origin, any geographic references in the quotation, and the outlet where the quote and article appeared. The result is an enriched dataset that allows us to look beyond the headlines of an article to understand the voices in each news item that are informing the content that European audiences consume. Our dataset has 618,328 quotes from 123,396 articles published between May 2018 and October 2019. It includes articles from 454 outlets labeled with one of 167 topics and 418 sentiments.

\subsection*{Quotation Matching}
Matching quotes is a nontrivial natural language processing task. Translation errors, paraphrasing, and excerpting lead to substantial differences between “identical” quotes (Vaucher et al., 2021). We implement a new method for identifying matching sentences by pairing a state-of-the-art transformer language model with density-based clustering. We begin by translating our quote corpus into an embedded space using a pre-trained SBERT model, a language model optimized for semantic textual similarity tasks (Reimers \& Gurevych, 2019). We then reduce the dimensionality of the embeddings by taking the first 70 principal components. Finally, we cluster the transformed embeddings using HDBSCAN such that quotes in the same cluster represent the same quote (Campello et al., 2013; McInnes et al., 2017). We capture 94\% of all matching quotes in the dataset while 78\% of the quotes we matched are correct matches. Of the 618,000 original quotes, 367,000 matched another quote in the dataset. Of these matched quotes, 200,000 were published by more than one outlet. More details on clustering performance and optimization, along with examples of matched and mismatched quotes, can be found in the supplemental materials.  

\subsection*{Saliency-Weighted Potential Quote Influence}

Our next task is to transform these matched quotations into signals of agenda setting, where potential influence is indicated by one outlet using the same quote as another outlet. The naïve approach would be to count all instances of quote matches where the labeled topic, sentiment, and speaker also match. We improve upon this approach by drawing on three heuristics that approximate when quote matches may be a signal of influence. First, p\textit{otential influence should only be attributed for quotes used after exposure.} There are instances where an outlet $j$ uses a quote both before and after exposure to outlet $i$'s initial quote use. It would not make sense to credit outlet $i$ with potential influence on outlet $j$'s use of the quote prior to exposure. Second, \textit{several outlets using the same quote suggests that the quote may be general knowledge}. A common example is a landmark statement by an important figure, such as President Trump’s official announcement of United States’ departure from the Paris Climate Accords. Many outlets quote such statements as basic information on the reported events. Such quotes provide weaker indication of influence between outlets and are discounted by their likelihood to be general knowledge. Finally, \textit{there is diminishing signal on the number of times an outlet uses a quote}. In other words, the first time an outlet uses a quote is more informative than the tenth time. We model both general knowledge and diminishing signal as the monotonically increasing and concave square root function. The following equation incorporates the above heuristics to transform a set of quotes $q$ into saliency-weighted potential quote influence of outlet $i$ on outlet $j$:

\begin{equation}
K_{ij} = \sum_q \frac{1}{\text{\# of outlets using }q} \times \sqrt{\text{\# of times }j\text{ uses }q} \times \frac{\text{\# of times }j\text{ uses }q\text{ after }i\text{'s first use}}{\text{\# of times }j\text{ uses }q}
	\label{eq:saliency_quote_influence}
\end{equation}

\noindent Inside the summation, the first term models the general knowledge discount, the second term models the diminishing signal on the number of times an outlet uses a quote, and the third term is the fraction of quotes after initial exposure. Further discussion can be found in the supplemental materials. Figure \ref{fig:quote_saliency_explainer} provides an example of how saliency-weighting is applied on a potential quote influence of \textit{AFP} on \textit{Sputnik}.

\begin{figure}
	\centering
	\includegraphics[width=\linewidth]{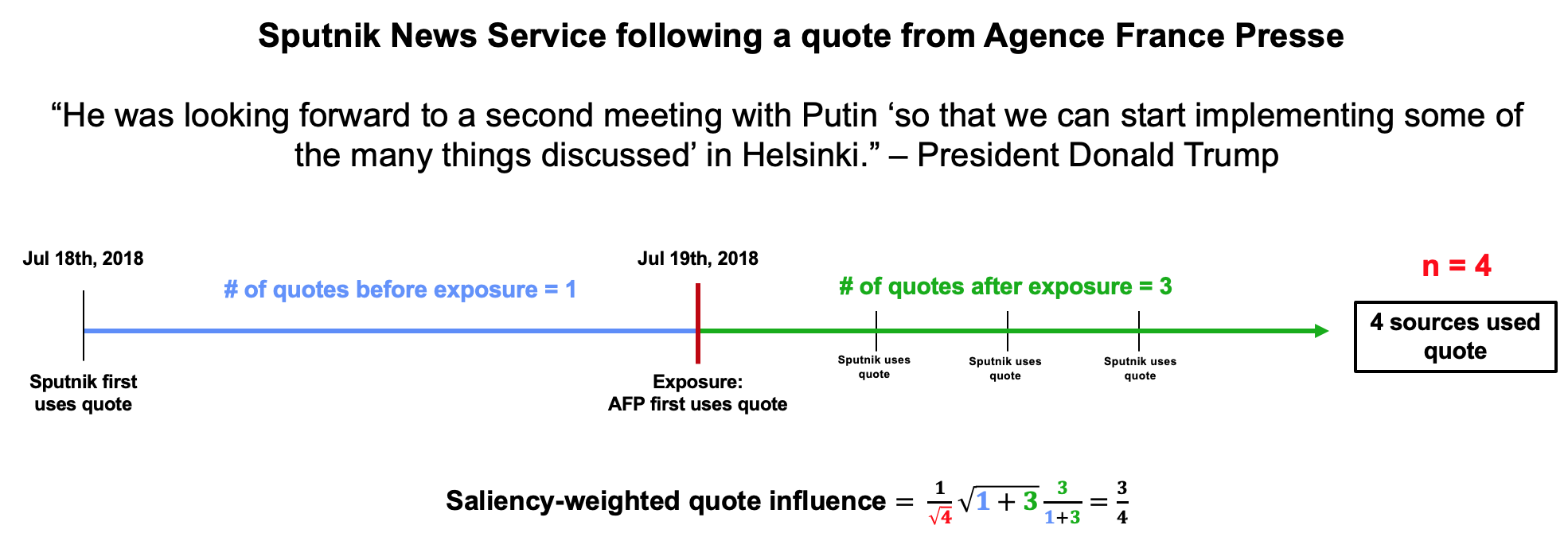}
	\caption{Saliency-weighted quote influence example. \textit{Sputnik} first used this quote on July 18th. \textit{AFP} first used it the next day on July 19th. \textit{Sputnik} then used the quote three more times. So, three out of four instances of \textit{Sputnik} using the quote came after exposure to \textit{AFP}. A total of four outlets used the quote. For this specific quote, \textit{AFP} has a saliency-weighted potential quote influence of $\frac{3}{4}$ on \textit{Sputnik}.}
	\label{fig:quote_saliency_explainer}
\end{figure}

\subsection*{Network Construction}

\begin{figure}
	\centering
	\includegraphics[width=\linewidth]{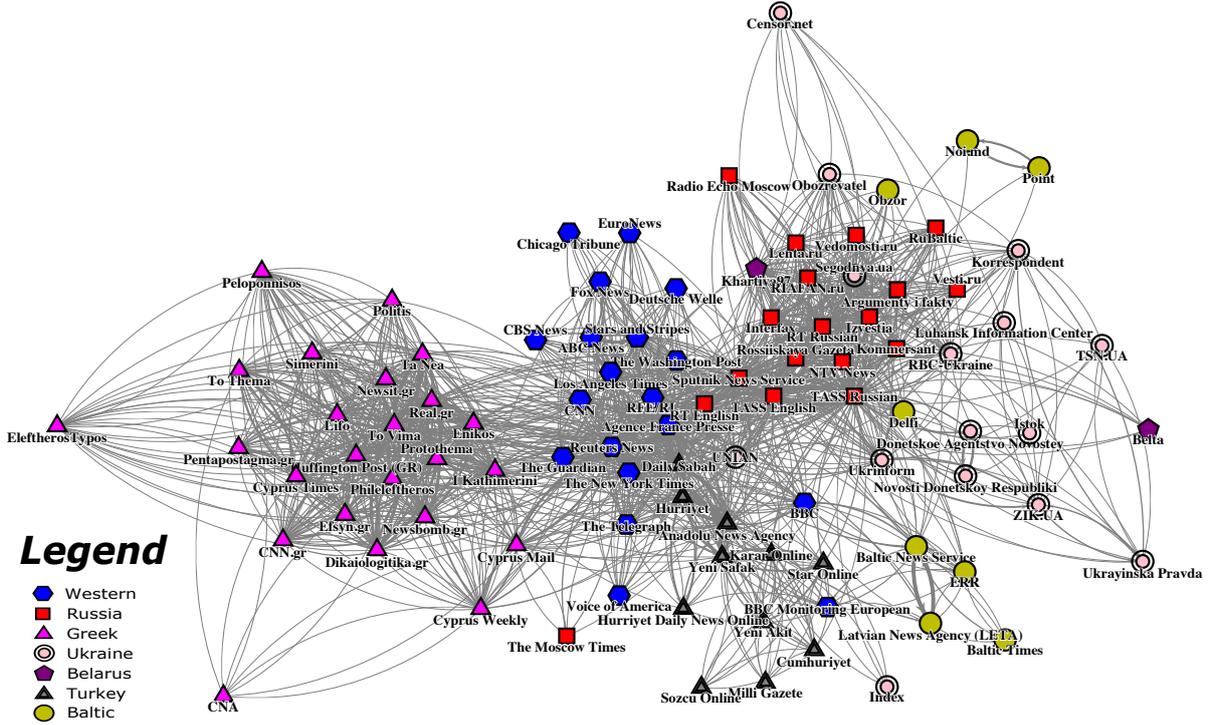}
	\caption{Constructed influence network among European media outlets. Edges represent past patterns of quote following with saliency weighting.}
	\label{fig:inf_network}
\end{figure}

We construct an influence network between media outlets using these saliency-weighted quote influences. Each node in the network represents a media outlet, with directed and weighted edges representing potential influences between them, represented by the adjacency matrix . Because actual influence is not directly observable, each network edge is modeled as a random variable with Poisson prior distribution parameterized by the observed saliency-weighted quote influence, , for realistic network interactions (Kao, 2019). Figure \ref{fig:inf_network} depicts the influence network constructed using quotes from all topics and sentiments with node positions determined by a force-directed graph drawing algorithm (Hu, 2005). Outlets generally cluster by geography, with outlets from the same country forming tightly connected communities. These communities are primarily connected to each other by wire services like \textit{Agence France Presse}, \textit{UNIAN}, \textit{Daily Sabah}, and \textit{Sputnik News Service}, which straddle geographic boundaries. Although not a wire service, the English-language version of the Russian outlet \textit{RT} (referred to as \textit{RT English} in the network) is also located along the spine connecting the eastern Europe media environment to the Western media environment. There are some exceptions to strict geographic clustering, which can be explained by language and outlet orientation. For example, \textit{The Moscow Times}, an English-language, independent media outlet, is separate from the main Russian community. The Russian outlet, \textit{Radio Echo Moscow}, is removed from the main cluster of Russian media outlets due to its liberal orientation but it nevertheless is closer to the cluster than \textit{The Moscow Times} given its Russian-language reporting.

\subsection*{Causal Impact Estimation}
The final stage of the methodology performs causal inference to estimate the causal impact between outlets using the saliency-weighted quote influence and the constructed influence network prior. This causal inference framework over networks was developed by Kao and Rubin (Kao, 2017; Imbens \& Rubin, 2015) and demonstrated on social media networks by Smith et al. (2018; 2021). This paper is its first demonstration on news media networks.  

The saliency-weighted quote influence only represents \textit{potential influence} between the outlets. Actual causal impact of outlet $i$ on outlet $j$ is explicitly captured in the causal estimand: 

\begin{equation}
	\label{eq:}
	\zeta_{i,j} \buildrel{\rm def}\over= Y_j (\bm{z}_{i+}, \bm{A}) - Y_j (\bm{z}_{i-}, \bm{A})
\end{equation}

\noindent where $Y_j (\bm{z}_{i+}, \bm{A})$ represents potential outcome on outlet $j$’s saliency-weighted quotes, with the binary vector $\bm{z}_{i+}$ denoting outlet $i$ being the quote source on the influence network represented by the directed and weighted adjacency matrix $\bm{A}$. Outlet $i$’s causal impact on outlet $j$ is established by subtracting the counterfactual outcome of outlet $j$’s saliency-weighted quotes in the absence of outlet $i$ as the source, $Y_j (\bm{z}_{i-}, \bm{A})$. In summary, the difference between these two potential outcomes represents outlet $i$’s causal impact on outlet $j$’s outcome. The first potential outcome $Y_j (\bm{z}_{i+}, \bm{A})$ is observed as outlet $i$’s saliency-weighted quote influence on outlet $j$,  in Equation \ref{eq:saliency_quote_influence}. However, the counterfactual outcome is not directly observable. It is imputed using a Poisson generalized linear mixed model (GLMM) fitted to the observed outcome on each outlet $j$. This outcome model captures the effects of each $n$-hop exposure to the quote source via the influence network, network confounders such as node degrees and node community membership, and heterogeneity between the outlets. Details of the model and inference can be found in the supplemental materials.

\section*{Results}
\subsection*{Causal Impact and Impact Slant}
Focusing our analysis on the impact of state-controlled outlets on the overall media landscape, we analyze Russian state-controlled media as the example and how it shapes media sentiment favorable towards its competition with the U.S. Specifically, we process the quote data with either a pro-Russia (i.e. positive to Russia or negative to the U.S.) or a pro-U.S. (i.e. positive to the U.S. or negative to Russia) sentiment. For each media outlet, both pro-Russia and pro-U.S. impact scores are estimated according to the causal impact it had on quotes of the corresponding sentiment. Total impact represents the sum of both pro-Russia and pro-U.S. impacts. In this paper, we define impact slant as the difference between an outlet’s impact in competing sentiments (e.g. pro-Russia versus pro-U.S.). Having to assign a directionality for visualizing results, we choose pro-Russia impact slant to be negative and pro-U.S. slant, positive. 

We begin by presenting results on the absolute, non-normalized causal impact to compare the magnitude of impact between outlets. Later analysis in Figure 8-10 is on the fraction of quote outcome caused by specific outlet sources, where causal impact is normalized by dividing by the maximum potential impact (i.e. the saliency-weighted quote influence). A normalized impact $x$ of outlet $i$ on outlet $j$ can be thought of as ``of the times $j$ uses the same quotes as $i$, $x\%$ of the time it was caused by exposure to $i$.''

\subsection*{Average Outlet Impact and Slant}
\begin{figure}
	\centering
	\includegraphics[width=.5\linewidth]{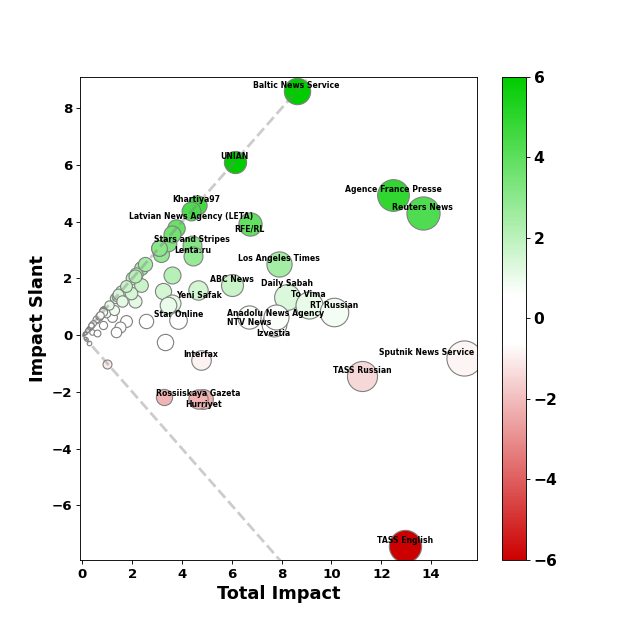}
	\caption{Media outlet impact slant and total impact across all topics. Outlets are sized by total impact and colored by impact slant. Neutral outlets are white. The more pro-U.S. or pro-Russia slanted an outlet is, the more green or red, respectively. The grey dotted lines indicate positions where all impact came in either the pro-U.S. or pro-Russia slant.}
	\label{fig:all_submessages_total_v_slant}
\end{figure}

In an analysis of the impact slant of outlets (pro-Russia vs. pro-U.S.), we find that, as expected, Western sources skew pro-U.S., while Russian sources skew pro-Russia. Figure \ref{fig:all_submessages_total_v_slant} summarizes each outlet’s average total impact and impact slant. The majority of outlets are neutral overall. A few outlets, like \textit{Baltic News Service}, \textit{UNIAN}, and \textit{Khartiya97}, exhibited only pro-U.S. impact, which suggests they almost exclusively publish quotes favorable to the U.S. Most of the high impact outlets have a mixture of pro-Russia and pro-U.S. impact.

Wire outlets like \textit{Sputnik News Service}, \textit{TASS}, \textit{Reuters News}, and \textit{Agence France Presse} have the highest impact level, which is consistent with their network centrality. It is also consistent with their increased article volume and role in “breaking” news stories. While impact slants largely match intuition, there are some surprises in the neutrality, and even slightly pro-U.S. slant, of certain Russian outlets. \textit{RT Russian}, for example, skews slightly pro-U.S. We investigate some of these specific outlets in the following section.

\subsection*{Examining Individual Outlet Impacts}
\begin{figure}
	\centering
	\hbox{\hspace{-3em} \includegraphics[width=1.15\linewidth]{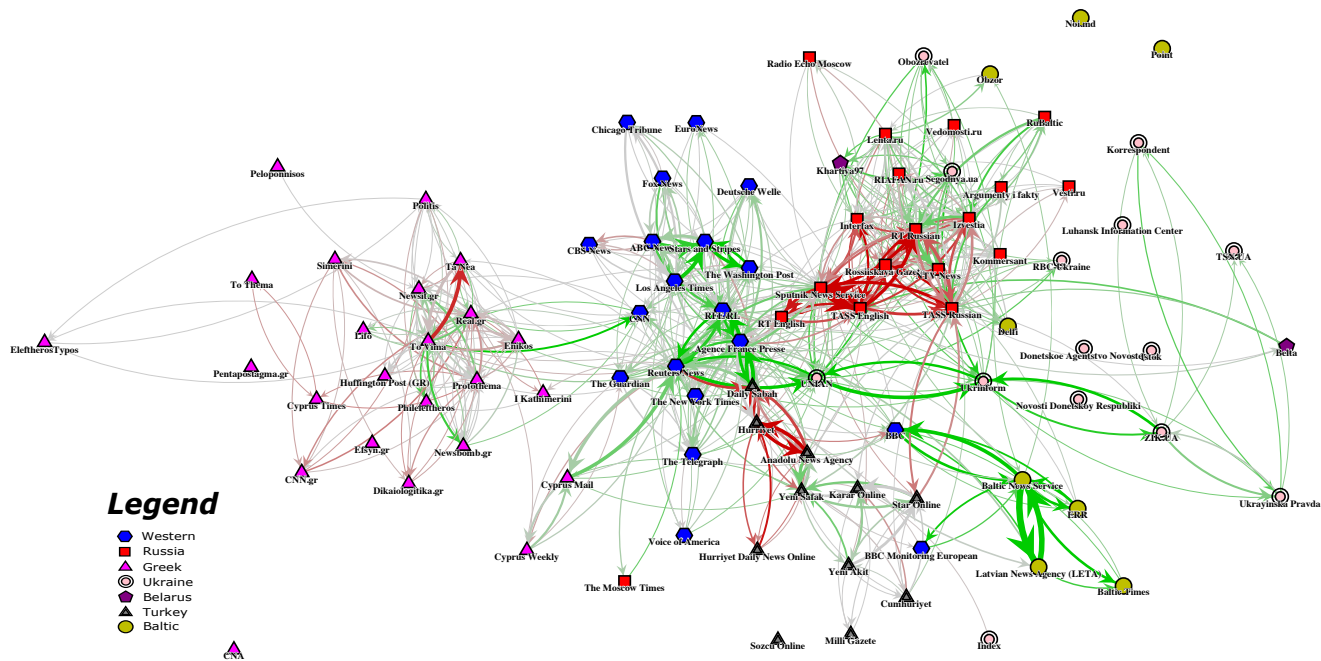}}
	\caption{Impact network across all topics. Edges are sized according to total impact and colored according to impact slant. Red indicates pro-Russia impact slant, green indicates pro-U.S. impact slant, and grey indicates neutral impact.}
	\label{fig:all_submessages_network_impact}
\end{figure}

\begin{figure}
	\centering
	\includegraphics[width=.9\linewidth]{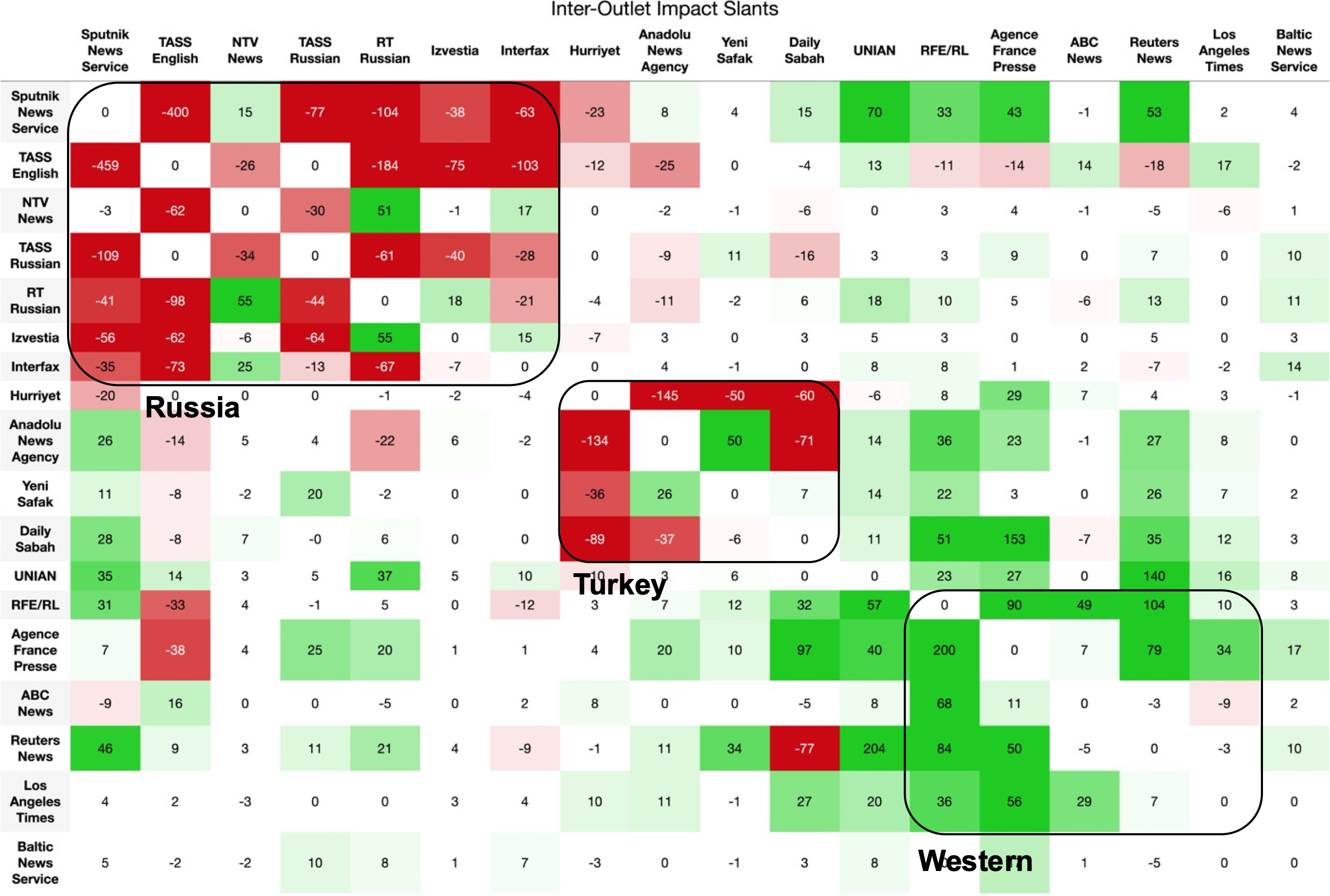}
	\includegraphics[width=.9\linewidth]{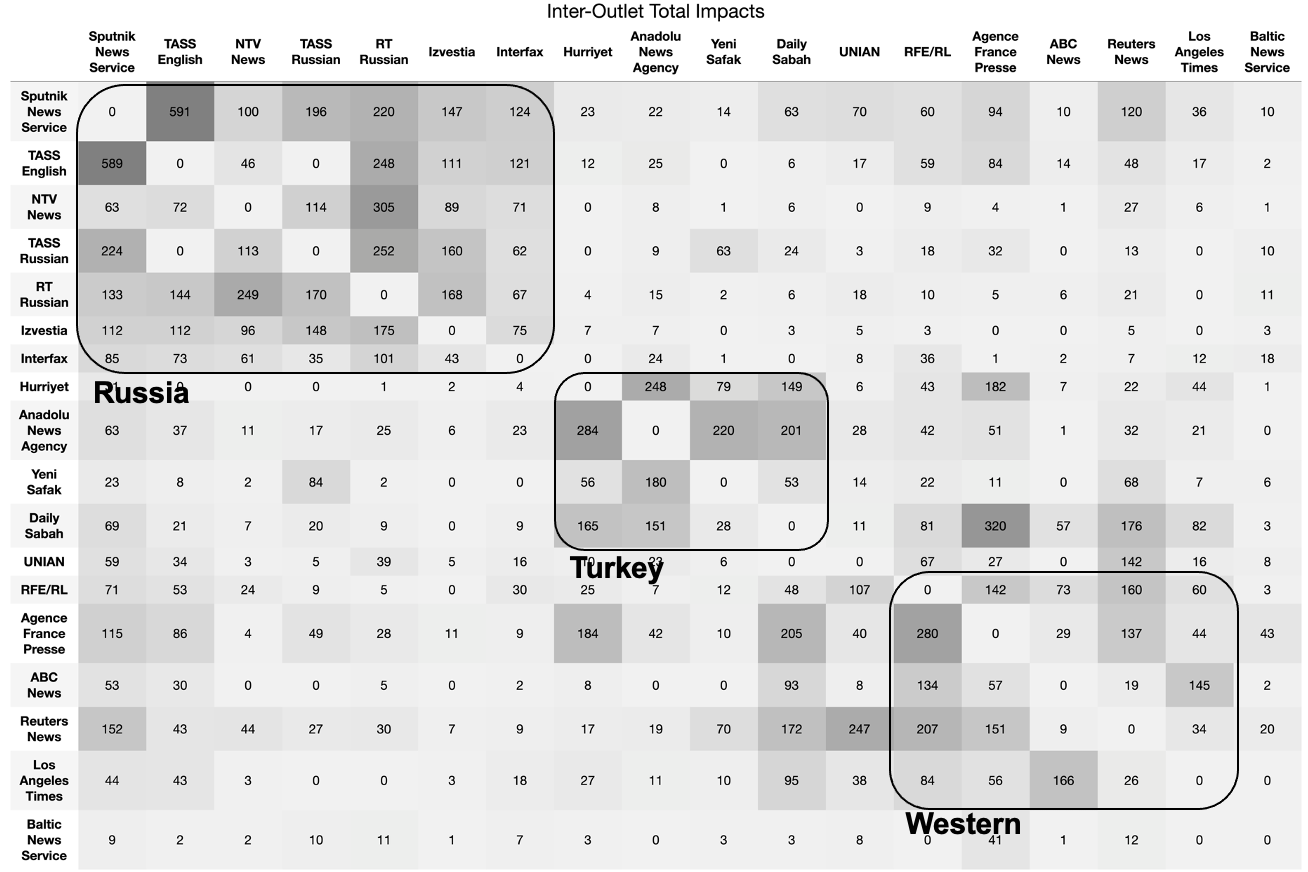}
	\caption{Impact slant (top) and total impact (bottom) of the 18 highest impact outlets. The row represents the source outlet and column represents the impacted outlet. In the top table, cells are colored according to their slant, with pro-Russia being red and negative, neutral white and around zero, and pro-U.S. green and positive. In the bottom table, cells are shaded according to the magnitude of total impact.}
	\label{fig:pairwise_impact_tables}
\end{figure}

Our methodology enables us to examine the impact of each outlet on other specific outlets. Figure \ref{fig:pairwise_impact_tables} summarizes the pairwise impacts of the most impactful outlets, while Figure \ref{fig:all_submessages_network_impact} displays impacts as edges on the impact network. Evident in the network community structure of the estimated causal impact between outlets, outlets typically influence others within their own geo-political media ecosystem.  This makes sense given language barriers between countries and the saliency of topics within a given country. Wire outlets, like \textit{Reuters}, \textit{AFP}, \textit{Sputnik} and \textit{TASS}, break this trend with significant impact reaching outside of their respective countries' media environments. These news wires also have the greatest total impact across U.S. and European media environments. Discovered as the notable bridges of the estimated causal impact network, these wire outlets play the role of connecting the media ecosystems of different countries. Nearly all the pro-Russia impact is between Russian outlets and Turkish outlets; outside of Russia and Turkey, net impact is largely pro-U.S. Moreover, the imbalance in the directionality of the impact slant shows that outlets selectively follow quotes. For example, Western outlets follow the pro-U.S. quotes from Russian outlet, while Russian outlets follow the pro-Russia quotes from Western outlets. 

Figure \ref{fig:all_submessages_network_impact} and Figure \ref{fig:pairwise_impact_tables} reveal a Russian echo chamber in which Russian state media outlets extensively borrow quotes from each other. Much of the net pro-Russia sentiment across European outlets comes from a few Russian outlets (\textit{TASS}, \textit{Sputnik}, \textit{RT}). The volume of quote following among these outlets is much higher than any other outlet pairs. This can signify attempts by Russian state media to promote their narratives to the Russian audiences. 

Figure \ref{fig:pairwise_impact_tables} also sheds light on \textit{RT Russian} and \textit{Izvestia}’s neutrality in impact slant. Compared to the other large Russian sources like \textit{Sputnik} and \textit{TASS}, \textit{RT Russian} and \textit{Izvestia} have much smaller amounts of influence. They do not seem to be as engaged in the Russian state media echo chamber. \textit{Izvestia} and \textit{RT Russian} even exert pro-U.S. net influences on each other, which can be attributed to the quotation of U.S. officials. Many of the most impactful quotes published by \textit{RT Russian} are from U.S. officials criticizing Russia. Due to the editorial context surrounding the extracted quotes, these U.S. statements that are classified as ``negative to Russia'' often reflect negatively on the U.S. to the audiences of Russian media within the context of the whole article. For instance, when \textit{RT Russian} quotes former U.S. Secretary of State Mike Pompeo saying that Russian violated the INF Treaty, commentary or a quote from the Kremlin follows, claiming the U.S. in fact violated the INF Treaty. Thus, as Russian media introduces some amount of U.S. statements for context setting purposes, audiences of Russian media may be conditioned to perceive U.S. statements as duplicitous or hypocritical. \textit{RT Russian}'s overall pro-U.S. impact in fact indicates that the outlet is a common source of U.S. official statements for other Russian outlets. This finding is supported by the fact that \textit{RT} has a well-established English-language service with access to press conferences or other English-language sources of U.S. official statements.

\begin{figure}
	\centering
	\includegraphics[width=\linewidth]{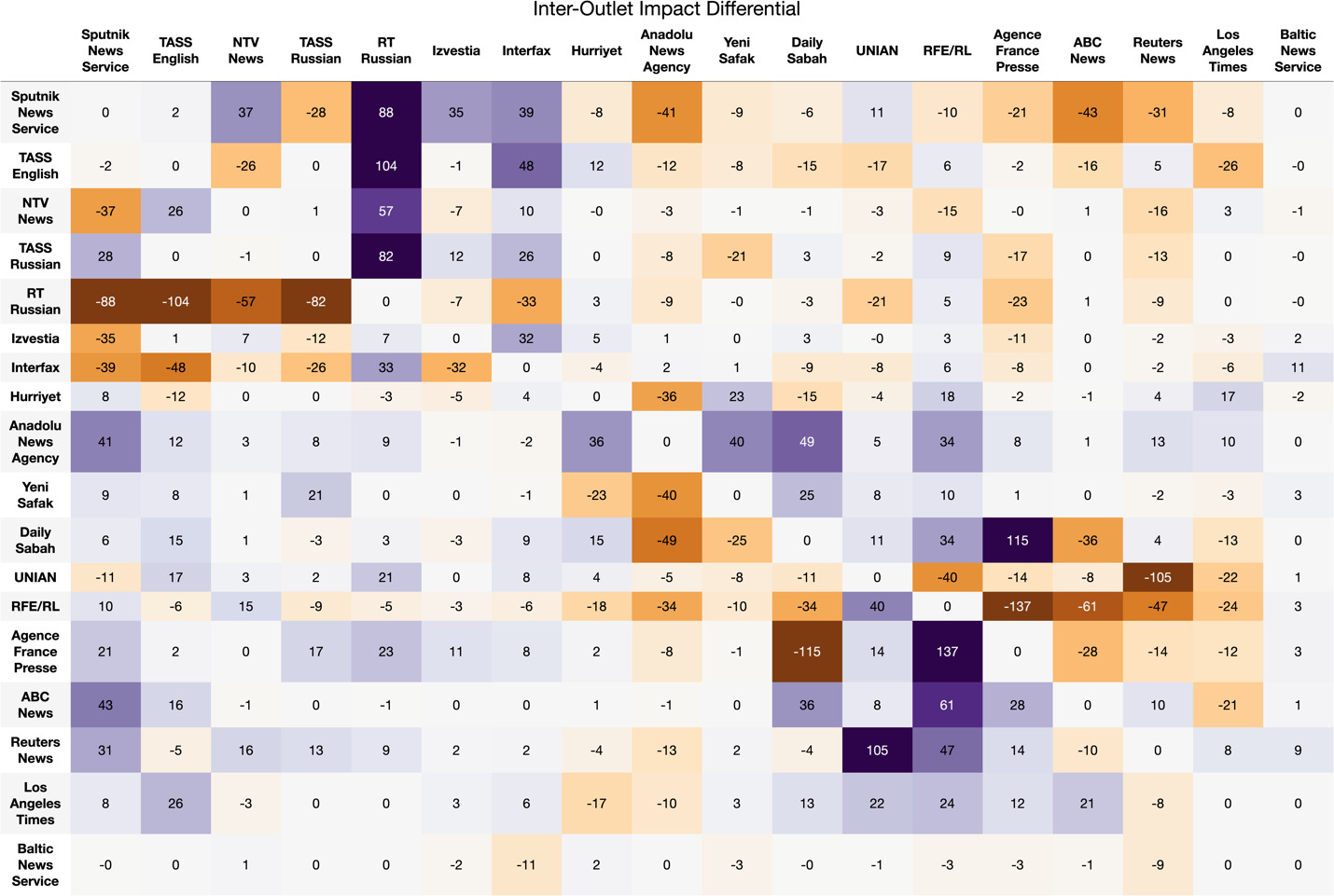}
	\caption{Total impact differentials of the top 19 outlets. These values represent the net impact of the row outlet on the column outlet. Negative, brown values mean the column outlet had more impact on the row outlet, while positive, purple values mean the row outlet had more impact than the column outlet. White, small magnitude values suggest symmetrical impact.}
	\label{fig:impact_differential}
\end{figure}

We can also examine the directionality of influence by looking at total impact differential (Figure \ref{fig:impact_differential}). An outlet $i$’s total impact differential with respect to outlet $j$ is defined by the impact of $i$ on $j$ minus the impact of $j$ on $i$. In general, impact is symmetric with outlet $i$ influencing outlet $j$ about as much as outlet $j$ influences outlet $i$; this leads to an impact differential near zero, as is the case between \textit{Sputnik} and \textit{TASS English}. However, there are some outlets that tend to have asymmetric influence. \textit{RT Russian} stands out as a follower, or an outlet that is influenced more than it influences. Looking at the row for \textit{RT Russian} in Figure \ref{fig:impact_differential}, many of the values in the Russian area are quite negative and brown, which indicates Russian outlets exert more influence on \textit{RT Russian} than \textit{RT Russian} exerts on them. Equivalently, the column for \textit{RT Russian} is positive and blue. In the other direction, some outlets are influencers and tend to influence more than they are influenced. One example of an influencer outlet might be \textit{Sputnik} or \textit{Anadolu News Agency}, who generally demonstrate net influence on other outlets in their media ecosystem. While \textit{Sputnik} is only an influencer in the Russian sphere, \textit{Anadolu News Agency}’s positive differential persists with non-Turkish outlets as well, although to a lesser extent. \textit{RFE/RL} is more often influenced by Western outlets such as \textit{Agence France Presse}, \textit{ABC News}, and \textit{Reuters News}. This tracks well with \textit{RFE/RL}'s mission of bringing Western-style reporting to other countries, not being a prominent outlet in the Western media environment. That Turkish outlet \textit{Daily Sabah} has high asymmetric influence over \textit{Agence France Presse} could suggest \textit{Agence France Presse} relies on \textit{Daily Sabah} in reporting on matters pertaining to Turkey.

\subsection*{Obscured Russian State-Affiliated Media Influence}
Our approach for intermedia agenda setting analysis reveals an obscured level of Russian influence, which would be of interest to any researchers of influence operations, especially on Russian media influence in Europe. Media analysts often categorize outlets by their political orientation for the purpose of understanding the expected sentiments in their content. These classifications are made based on a qualitative assessment of an outlet's published content along with ownership analysis. Of great importance in the studying of the Russian media environment is knowing whether the outlet is state-controlled, state-agenda, or independent. This classification determines to what extent the Russian state is controls the content being published in an outlet. State-agenda outlets, for instance, are categorized as not being directly controlled by the state, but as advancing the state agenda due to ownership or political ties to the state. This type of Russian state media influence is rather overt and can usually only be masked by complex ownership schemes. 

\begin{figure}
	\centering
	\includegraphics[width=.7\linewidth]{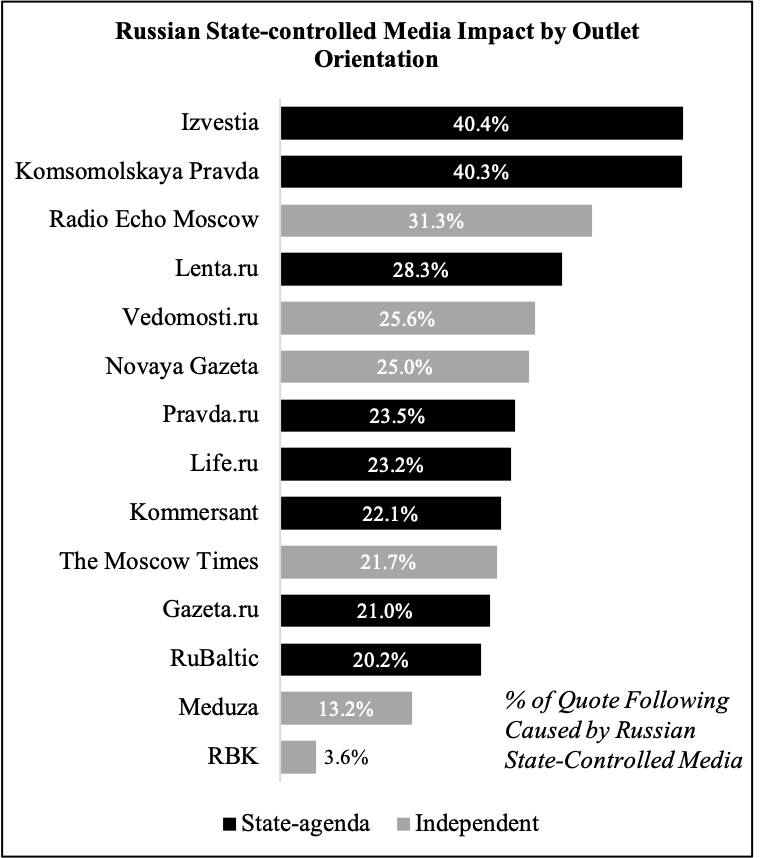}
	\caption{Russian state-controlled media impact on other Russian outlets by traditional orientation classification}
	\label{fig:novetta-1}.
\end{figure}

Our approach shows on a very detailed level to what extent other outlets are reporting within the parameters set by Russian state-controlled media. As shown in Figure \ref{fig:novetta-1}, Russian state-controlled media have an average causal impact on 24.2\% of quote following by other Russian outlets. This is important to know because while an outlet may be independent and report in liberal terms, they nevertheless report on the same topics and use the same sources as Russian state media. \textit{Radio Echo Moscow} exemplifies this condition, having 31.3\% of its quote following impacted by Russian state media. This outlet has previously been noted for its independent from the Russian state and often reports in liberal and pro-Western terms (Oates \& McCormack, 2010). However, as shown in Figure \ref{fig:novetta-1}, this outlet was measured to frequently have Russian state-controlled media causing the use of identical quotes in its reporting. In other words, \textit{Radio Echo Moscow} is frequently reporting using the topics and sources Russian state-controlled media used first. While \textit{Radio Echo Moscow} is free from reporting in pro-Russian terms, the outlet is not free from reporting the agenda Russian state-controlled media has set.

\subsection*{Media Independence as a Limit on Russian State Media Influence}
By dividing media outlets in each European country into state-affiliated (e.g. state-controlled and state-agenda) and independent, one can see in Figure 9 that on average Russian state media has greater influence on state-affiliated media than on independent media in 11 out of 12 European countries with monitored state and independent outlets. This is possibly due to the simplicity of relying on the packaged reporting from news wires like Russian state-controlled \textit{TASS} in countries that do not have news wires with such extensive sources. These results may also witness of Russia's political dominance in that the state-controlled media of various countries are reacting to topics of interest to Russia. Regardless of the possible causes, the results of our analysis suggest media independence may decrease the impact that Russian state media has on the European media environment. Only Ukrainian media bucks the trends observed in the rest of European media. That Ukrainian state-controlled outlets have less reliance on Russian reporting than do Ukrainian independent outlets could be a reflection of Ukraine’s foreign policy being oriented toward the West away from Russia since 2014 (Shyrokykh, 2018; Dragneva \& Wolczuk, 2016; Boulègue \& Litra, 2019). 

\begin{figure}
	\centering
	\includegraphics[width=.7\linewidth]{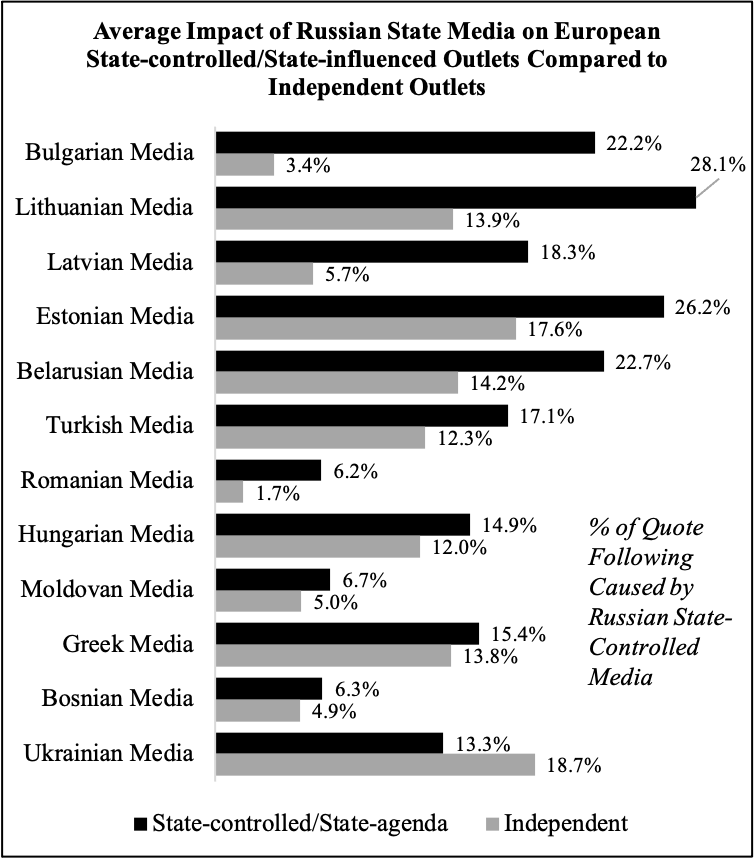}
	\caption{Average impact of Russian state media on state-controlled/state-agenda media vs. independent media in Europe.}
	\label{fig:novetta-2}
\end{figure}

\begin{figure}[t!]
	\centering
	\includegraphics[width=.708\linewidth]{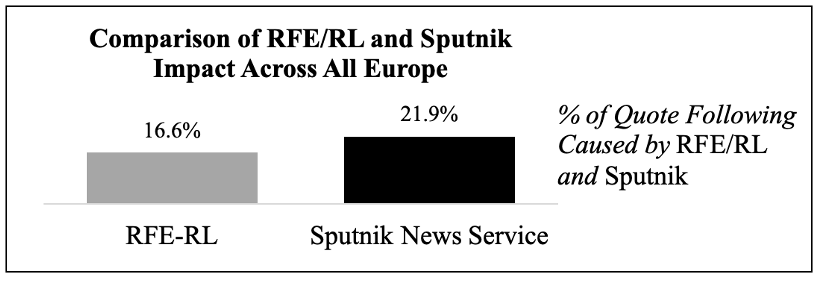}
	\includegraphics[width=.7\linewidth]{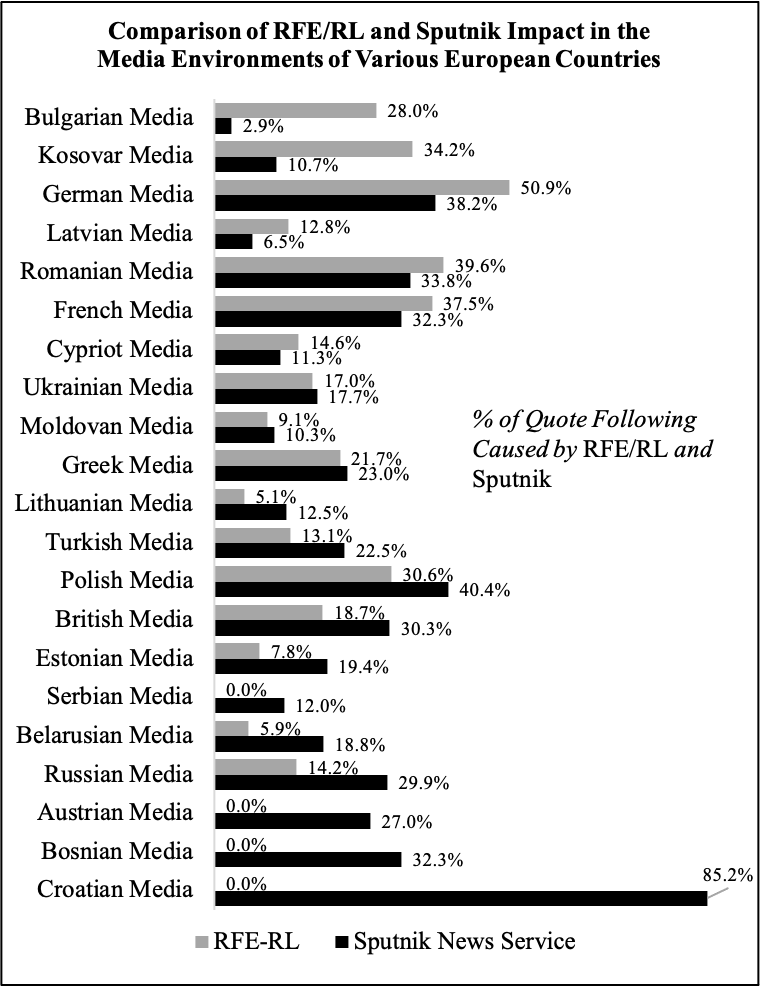}
	\caption{Comparison of \textit{RFE/RL} and \textit{Sputnik News Service} impact in the media environments of various European countries.}
	\label{fig:novetta-3}
\end{figure}

\subsection*{The Impact of U.S. and Russian Government-Sponsored Media in Europe}
Comparing the reach of the two state-sponsored media, the Russia-sponsored \textit{Sputnik} has greater overall impact on European media than does the U.S.-sponsored \textit{RFE/RL}, as shown in Figure \ref{fig:novetta-3}. \textit{RFE/RL} nevertheless has an advantage in the media of the following countries: Bulgaria, Kosovo, Germany, Latvia, Romania, France, and Cyprus. Russian information and influence campaigns are more commonly referenced in the Baltics, the Balkans, and in Eastern Europe. These findings provide a quantifiable measure of how much impact one of Russia's influence tools is having in those regions. 

\subsection*{By Specific Topic: The End of the Intermediate-Range Nuclear Forces Treaty in 2019}
One advantage of our approach is the ability to map networks of narrative dissemination and impact slant for any specific topic or range of sentiments within our data's established taxonomy. We have already demonstrated the sentiment-based result generation of pro-Russia versus pro-U.S. in the above results based on all topics. The data could also be used to measure impact on facets beyond sentiment, such as determining whether outlets quote Western officials more frequently than Russian officials. 

Using our method to home in on specific coverage topics also presents an alternate approach for answering the question posed by Stern et al. (2020) on whether "bellwether" outlets—ones that influence other outlets' coverage of a certain topic—can be detected. Below, we focus our quote analysis on the topic on the end of the Intermediate-Range Nuclear Forces (INF) Treaty in 2019. To avoid unnecessary confounding correlation between the network and the quote outcomes, we construct the influence network using all quotes except those on the nuclear treaty topic. Due to smaller data size relative to analysis on all topics, the estimated impact has higher uncertainty. However, the result below still matches with intuition and maps out the influential outlets in shaping the sentiments on this specific topic.

\begin{figure}[h]
	\centering
	\includegraphics[width=.5\linewidth]{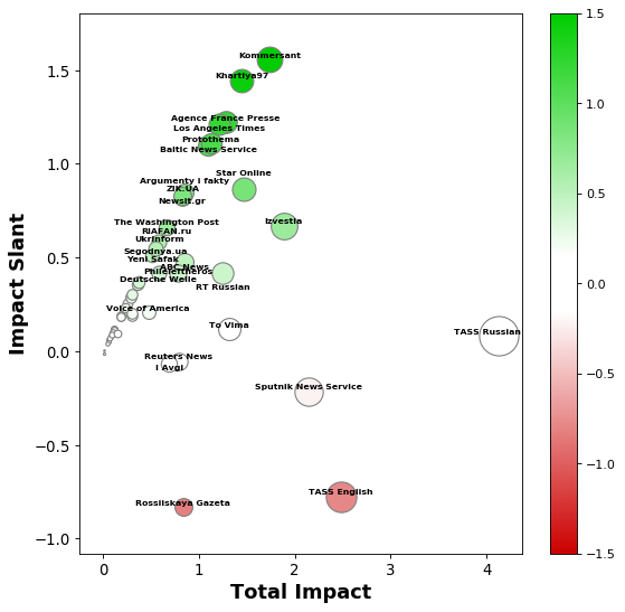}
	\caption{Media outlet impact slant and total impact on the nuclear treaty topic. Outlets are sized by total impact and colored by impact slant. Neutral outlets are white. The more pro-U.S. or pro-Russia slanted an outlet is, the more green or red, respectively. The grey dotted lines indicate positions where all impact came in either the pro-U.S. or pro-Russia slant.}
	\label{fig:nuclear_coop}
\end{figure}

\begin{figure}
	\centering
	\includegraphics[width=1\linewidth]{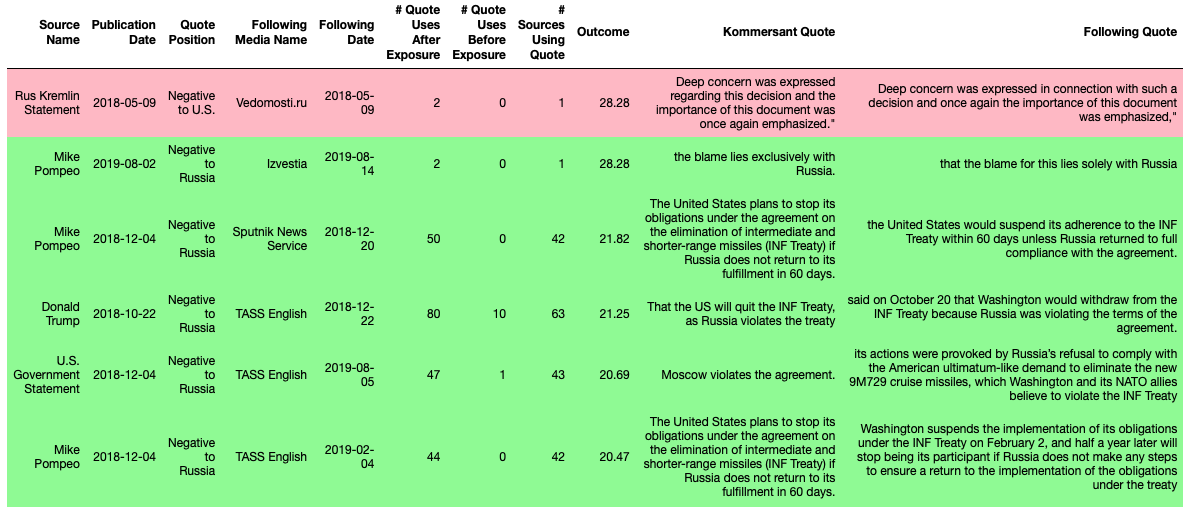}
	\caption{Examples of top followed quotes for \textit{Kommersant}. Many of the top quotes were negative to Russia, which explains why \textit{Kommersant} was pro-U.S. over the nuclear treaty topic.}
	\label{fig:kommersant}
\end{figure}

From the results on the nuclear treaty topic, one can see in Figure \ref{fig:nuclear_coop} that \textit{Rossiiskaya Gazeta} and \textit{TASS English} had the most pro-Russia impact on other outlets. In other words, these outlets published quotes codified as "positive to Russia" and "negative to U.S." most frequently in European and U.S. media. \textit{TASS Russian} had the highest impact on the topic of nuclear treaty, meaning it was a significant determiner of the direction coverage of nuclear treaty took. Strikingly, \textit{TASS Russian's} impact was rather neutral, which again is due to other outlets sourcing from \textit{TASS Russian} for U.S. official statements on nuclear treaty with Russia. 

Our approach allows us to investigate the individual quotes that determined a specific outlet’s influence. For example, \textit{Kommersant} is pro-U.S., despite being Russian state-agenda media.  If we inspect \textit{Kommersant’s} top followed quotes in Figure \ref{fig:kommersant}, we see that they are mostly pro-U.S. \textit{Kommersant} had 3x as much pro-U.S. outcome as pro-Russia outcome despite publishing roughly the same number of pro-U.S. quotes as pro-U.S. quotes. Upon closer examination, \textit{Kommersant}’s pro-Russia quotes were not picked up by other Russian outlets, which limited \textit{Kommersant}’s pro-Russia impact. This is also consistent with \textit{Kommersant}’s network position on the periphery of the Russian state media community in Figure \ref{fig:inf_network}. 

\section*{Discussion and Conclusion}
In developing a quantitative approach to measure agenda setting between outlets by both topic and sentiment, we have identified a method to measure the impact that state-controlled outlets have on outlets that are not state-controlled. Using quotations as the unit of analysis, we have shown how Russian state-controlled outlets can create an echo chamber, determining how a topic is covered across the media ecosystem. Our approach also provides quantitative analysis to better understand patterns in media outlets’ selection of quotations and portrayal of sentiments when covering specific topics. This metric of impact also reveals an implicit, obscured level of influence where we can detect the extent that Russian state media sets the parameters for how other outlets report on a topic, regardless of the ownership or declared orientation (e.g. independent, liberal, opposition, etc.) of those outlets. Our results also identify those outlets that serve as bellwethers for particular topics, and those that serve as other outlets’ sources of quotations from particular types of speakers.

The implications of these findings on strategic communications and influence operations are significant. From a monitoring and evaluation standpoint, this approach and resulting metrics of impact can show the effectiveness of communications campaigns. It can also help independent observers or European governments track the relative performance and influence of outlets like Sputnik or RT on a European country’s media environment. Editorial boards striving for independence can measure the extent to which their coverage – by topic – is determined by the agenda set from state-controlled media. Such a metric is also of general use to consumers of information who want to know the extent to which a selected source of news or its coverage of a particular topic is influenced by state-controlled media.

For strategic communicators and those studying influence operations alike, this approach offers metrics of performance and effectiveness. Understanding which outlets have impact on a topic-by-topic basis can help in identifying key events for communication such as interviews; for those attempting to understand particular narratives or track disinformation on a topic, this method offers a quantitative approach to estimating causal impact on the broader information ecosystem.

\newpage
\glsaddall
\printglossary[nonumberlist]
\newpage
\input{manual_bib}

\newpage
\input{supplement}

\end{document}

%% file: manual_bib.tex
\def\bibindent{2em}

%% file: supplement.tex
\section*{Supplemental Materials}

\section*{Methodology}

Below are additional details on the methods used to automate quote matching and to estimate causal impact between outlets. 
\vspace{-0.35cm}
\subsection*{Quote Clustering}

\begin{figure}[h]
	\centering
	\includegraphics[width=.4\linewidth]{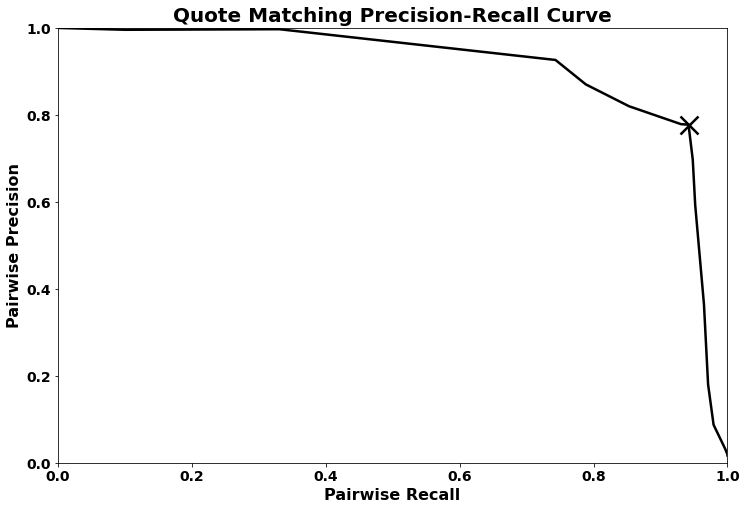}
	\vspace{-0.5cm}
	\caption{Precision recall curve for quote matching clustering. At the X (our optimum), we achieve 78\% precision and 94\% recall. That is, 78\% of identified quote matches are true matches and we capture approximately 94\% of all quote matches.}
	\label{fig:pr_curve}
\end{figure}

We optimized and tested HDBSCAN parameters using a hand-labeled truth set of 1668 quotes. Results from our clustering optimization can be seen in Figure \ref{fig:pr_curve}, with specific examples provided in Table \ref{table:quote_types}. At the optimal HDBSCAN parameters, we achieved 94\% recall and 78\% precision on our task of clustering quotes. In general, false positives tend to occur with quotes that use jargon, such as the quote discussing S-400s and F-35s. They also occur with subjects where there are not many quotes; these quotes may end up evenly-spaced in the embedded space and the density-based clustering will group them together. False negatives tend to occur on difficult paraphrase matching, or in areas where the same quote is used in slightly different form many times and the clustering algorithm fails to merge two nearby clusters. 

While there are no directly comparable benchmarks in the literature, we can compare our results to ``paraphrase identification'' results. Wang et al. report 89\% basic accuracy on paraphrase identification, where accuracy is the sum of true positives and true negatives divided by all labels. We demonstrate 99.4\% accuracy on our task. Admittedly, our quote matching task is less challenging than paraphrasing; while some of quote matching is paraphrase identification, many of the quotes are near-exact matches. Mellace et al. proposed a similar sentence matching pipeline using SBERT and DBSCAN; they analyze novels and cluster sentences that captured similar relations between characters, such as characters speaking to or smiling at other characters. (Mellace et al., 2020)

\begin{table}
	\scriptsize
	\centering
	\resizebox{\textwidth}{!}{
		\input{quote_matches.tex}
	}
	\caption{Examples of quote matches, successful and unsuccessful. True positives were quotes that matched correctly. False positives are quotes that matched, but should not have because they are actually different quotes. False negatives are quotes that did not match, but should have because they are the same quote. }
	\label{table:quote_types}
\end{table}

\vspace{-0.65cm}
\subsection*{Saliency-weighted Quote Influence Heuristics}

We apply three heuristics to transform sets of matched quotes into saliency-weighted quote influence. This section explains our assumptions more precisely. First, \textit{potential influence should only be attributed for quotes used after exposure}. We apply this heuristic by scaling our value by the proportion of matched quotes that came after possible exposure. Second, \textit{several outlets using the same quote suggests that the quote may be general knowledge}. We assume that the more outlets using the quote, the likelier it is to be general knowledge and the less signal the instance of quote following contains. Thus, we discount by some function of the number of outlets that used the quote. We represent this heuristic by function $g_1(x)$, a monotonically increasing and concave function. The final heuristic is \textit{diminishing signal on the number of times an outlet uses a quote}. More plainly, the first time an outlet uses a quote is more informative than the tenth time. We model this heuristic as a monotonically increasing and concave function, $g_2(x)$. Let $Q$ be the set of quotes outlet $i$ used, $S_q$ be the number of outlets that used quote $q$ and let $T_{q,i}$ be the time outlet $j$ used quote $q$. The following equation incorporates the above heuristics to transform the sets of matched quotes into saliency-weighted quote influence of outlet $i$ on outlet $j$:

\begin{equation}
	\kappa_{ij} = \sum_{q \in Q_i \cap Q_j} \frac{1}{g_1(S_q)} g_2(|q \in Q_j|) \frac{\left| \{ q \in Q_j \ : \ t_{q,i} \geq \min (t_{q}) \} \right|}{\sqrt{|q \in Q_j|}}
\end{equation}

Inside the summation, the first term models the general knowledge discount, the second term models the diminishing signal on the number of times an outlet uses a quote, and the third term is the fraction of quotes after initial exposure. For this analysis, we substitute the square root function as our monotonically increasing and concave function ($g_1(x) = g_2(x) = \sqrt{x}$), which gives us the following equation:

\begin{equation}
	\kappa_{ij} = \sum_{q \in Q_i \cap Q_j} \frac{1}{\sqrt{S_q}} \sqrt{|q \in Q_j|} \frac{\left| \{ q \in Q_j \ : \ t_{q,i} \geq \min (t_{q}) \} \right|}{\sqrt{|q \in Q_j|}}
	\label{eq:saliency_outcome}
\end{equation}

\subsection*{Potential Outcome Model for Causal Impact Estimation}

For causal impact estimation, the counterfactual outcomes $Y_j (\bm{z}_{i-}, \bm{A})$ are not directly observable. They are imputed using a Poisson generalized linear mixed model (GLMM) fitted to the observed outcome $Y_j (\bm{z}_{i+}, \bm{A})$on each outlet $j$. The Poisson distribution models the saliency-weighted quote outcomes as Poisson processes. Practically, the saliency-weighting makes the observed outcomes non-discrete so they need to be rounded before being fitted to the model. To reduce the rounding effect and increase the resolution of the Poisson-distributed outcome, the observed outcomes are multiplied by a scalar of ten before rounding. This makes a unit of causal impact in this paper equal to 0.1 saliency-weighted quote defined in Equation \ref{eq:saliency_outcome}. The Poisson GLMM uses the canonical log-link function and includes linear predictor coefficients ($\tau, \gamma, \beta, \mu$), corresponding to the source indicator $z_j$, n-hop exposures $s_j^{(n)}$, the covariate vector $\bm{x_j}$, and the baseline outcome:

\begin{equation}
\begin{gathered}
	Y_j(\bm{z}, \bm{A}) \sim \text{Poisson}(\lambda_j)\\
	\log(\gamma_j) = \tau z_j + \sum_{n=1}^{N_\text{hop}} s_j^{(n)} \tau \prod_{k=1}^{n} \gamma_k + \bm{\beta}^T \bm{x}_j + \mu + \epsilon_j
\end{gathered}
\end{equation}

In the five effect terms, $\tau z_j$  represents the primary effect of the quote source,  \\$\sum_{n=1}^{N_\text{hop}} s_j^{(n)} \tau \prod_{k=1}^{n} \gamma_k$ represents the accumulative network influence effect from each n-hop exposures  $s_j^{(n)}$ to the source, $\gamma_k$ (between 0 and 1) represents how quickly the effect decays over each additional $k$th hop, $\bm{\beta}^T \bm{x}_j$ is the effect of the unit covariates $x_j$ of network confounders including node degrees and node community membership (fitting the influence network to a blockmodel with 10 communities). These confounders are accounted for through covariate adjustment to disentangle actual causal impact from effects of homophily (birds of a feather flock together) and vertex degrees. Lastly, $\mu$ is the baseline effect on each outlet, and $\epsilon_j \sim \text{Normal}(0, \sigma^2_\epsilon = .1)$ provides independent and identically distributed variation for heterogeneity between the outlets. The amounts of social exposure at the nth hop are determined by $\bm{A}^{T^n} \bm{z}$. This captures quote diffusion via all exposure paths to the quote source over the influence network. Diminishing return of additional exposures is modeled using (elementwise) log-exposure, $s^{(n)} = \log (\bm{A}^{T^n}\bm{z} + 1)$.To account for the uncertainty of the influence network $\bm{A}$, it is jointly estimated with the model parameters ($\tau, \gamma, \beta, \mu$) through Markov Chain Monte Carlo (MCMC) and Bayesian regression.

\section*{Relationship Between Volume and Impact}
\begin{figure}
	\centering
	\includegraphics[width=.45\linewidth]{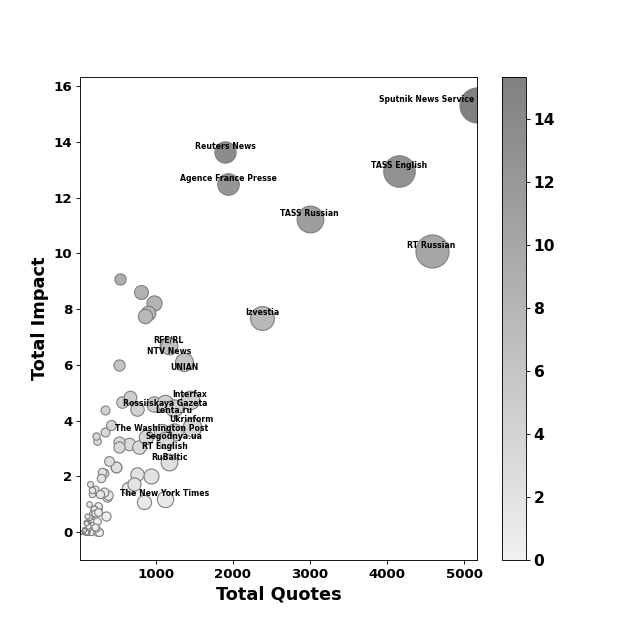}
	\includegraphics[width=.45\linewidth]{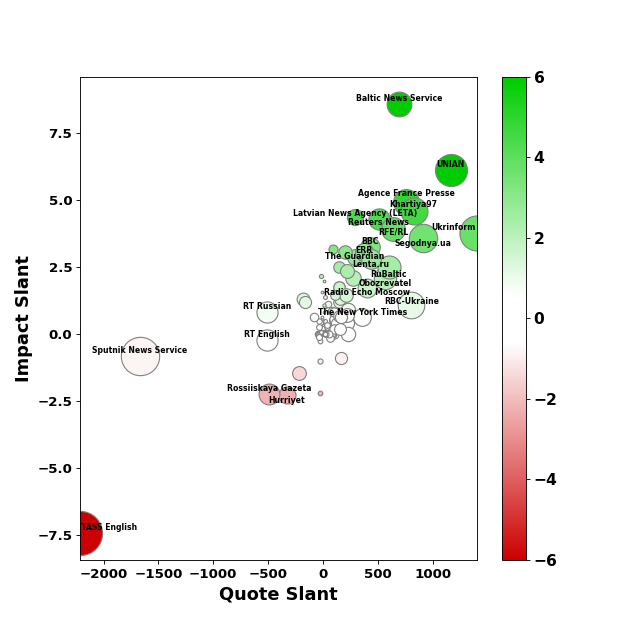}
	\vspace{-0.8cm}
	\caption{Relationship between quotes and impact. The left figure shows the relationship between total quotes and impact, while the right figure shows the relationship between quote slant and impact.  Points are sized according to the x axis and colored according to the y axis.  Pro-Russia slant is indicated as negative values and pro-U.S. positive.}
	\label{fig:quotes_and_impact}
\end{figure}

One relevant question considers the relationship between quotes and impact: do outlets that publish frequently have the most impact? Moreover, what is the connection between quote slant (how many pro-U.S. quotes you publish versus pro-Russia quotes) and impact slant? Figure \ref{fig:quotes_and_impact} examines this relationship. The left plot demonstrates that outlets that published more quotes were generally more impactful. \textit{Sputnik News Service}, the most impact outlet, also published the most quotes. However, the correlation is not perfect. Consider \textit{RT Russian}, which had over twice as many quotes as \textit{Reuters News} and \textit{Agence France Presse}, yet had less impact. There is considerable variation in total impact among outlets with the same number of quotes, demonstrating the role that causal inference on networks play to more accurately quantify actual impact from simple activity count statistics. 

A similar story holds when we compare impact slant to quote slant. There is a linear relationship between the two, but there is significant variation in impact slant among outlets with similar quote slants. \textit{Sputnik News Service}, for example, is fairly neutral despite very pro-Russia quote slant, which can be explained by its pro-U.S. quotes being generally more impactful and propagated by Western media outlets, shown in Figure 6 of the main paper. 

%% file: quote_matches.tex
\newcolumntype{L}[1]{>{\raggedright\arraybackslash}m{#1}}
\newcolumntype{C}[1]{>{\centering\arraybackslash}m{#1}}
\newcolumntype{R}[1]{>{\raggedleft\arraybackslash}m{#1}}
\renewcommand{\arraystretch}{1.7}
\begin{tabular}{C{1.5cm}|C{1.5cm} C{5cm} C{5cm}}
	\toprule 
	& \textbf{Source}                    & \textbf{Quote 1} & \textbf{Quote 2} \\ \midrule 
	\multirow{3}{*}{\textbf{\shortstack{True\\Positives}}}  & \shortstack{Russian\\Embassy\\Statement} &  Moscow "repeatedly answered the propaganda" calls of the "West" to release those under investigation, defendants or convicts in accordance with Russian law for serious crimes. 
	
	 \textit{RT Russian}, 2018-06-19    &      Russia has repeatedly and unequivocally responded to propaganda “calls” from the West for the release of certain suspects, defendants or convicts in accordance with our legislation for various serious crimes.

	\textit{TASS Russian}, 2018-06-17   \\
	& \shortstack{Maxim\\Oreshkin}            &     There is road-building machinery and a number of other items that Russia imports.
	
	\textit{Reuters News}, 2018-06-19    &       Among the goods in respect of which these duties may be introduced, "road construction equipment and a number of other elements that Russia imports." 

	\textit{Interfax}, 2018-6-20   \\
	& \shortstack{U.S.\\Government\\Statement} &     Russia's 9M729 missile violates the INF Treaty and Moscow's assurances that the missile's range is within the limits allowed by the agreement.
	
	\textit{Sputnik News Service}, 2019-06-28    &      Moscow in violation of the contract due to the presence of the 9M729 missile, the flight range of which allegedly exceeds acceptable standards.
	
	\textit{Izvestia}, 2019-07-26   \\ \hline 
	\multirow{2}{*}{\textbf{\shortstack{False\\Positives}}} & \shortstack{Hulusi\\Akar}               &   Works are under way to protect and protect the rights arising from international law and agreements, and that they will do their best to protect the rights of the blue citizens.
	
	\textit{Karar Online}, 2018-12-20      &     When we talk about Blue Homeland, we mean a 462km area that includes both the sea and the sky. We have rights set out in international treaties and expect our interlocutors and neighbors to respect them. In this context, we expect to move forward.
	
	\textit{To Vima}, 2019-08-12
	\\
	& \shortstack{Unnamed\\U.S.\\Officials}    &     Suggested Turkey buy the U.S. Patriot missile system rather than the S-400, arguing it is incompatible with NATO systems and is a threat to the F-35 fifth-generation stealth aircraft.
	
	\textit{Daily Sabah}, 2019-05-03    &     The government of US President Donald Trump still intends to impose sanctions on Turkey and exclude it from the F-35 construction program if it finally purchases the S-400.
	
	\textit{Protothema}, 2019-07-04    \\ \hline
	\multirow{2}{*}{\textbf{\shortstack{False\\Negatives}}} & \shortstack{Mike\\Pompeo}              &    The United States calls on Russia to respect the principles to which it has long claimed to adhere and to end its occupation of Crimea.
	
	\textit{CNN}, 2018-07-25     &      The U.S. would hold to its long-standing principle of refusing to recognize Kremlin claims of sovereignty over territory seized by force in violation of international law. And he called for Russia to respect principles it claims to respect and "end its occupation of Crimea.
	
	\textit{Los Angeles Times}, 2018-07-25   \\
	& \shortstack{Mevlut\\Cavusoglu}          &    We must first clarify what we will negotiate at the informal five-day conference and then supplement the content of the terms of reference.
	
	\textit{To Thema}, 2019-09-15     &     Stated that, for this reason, there is a need for a five-day informal conference, so that the sides can clarify what will be discussed next in the Cyprus issue. Then the terms of reference will be clear.
	
	\textit{I Kathimerini}, 2019-09-09    \\
	\bottomrule
\end{tabular}